\documentclass[%
 reprint,
 amsmath,amssymb,
 aps,
]{revtex4-2}
\usepackage{graphicx}
\usepackage{xspace}
\usepackage{amsmath}
\usepackage{amsfonts}
\usepackage{amssymb}
\usepackage{enumerate}
\usepackage{epsfig}
\usepackage{lineno}
\usepackage{xcolor}

\newcommand{\ksn}{\ensuremath{\mathrm{K_S}}\xspace}
\newcommand{\ksln}{\ensuremath{\mathrm{K_{S,L}}}\xspace}
\newcommand{\kln}{\ensuremath{\mathrm{K_L}}\xspace}

\newcommand{\kn}{\ensuremath{\mathrm{K^0}}\xspace}
\newcommand{\knb}{\ensuremath{\mathrm{\bar{K}^0}}\xspace}

\newcommand{\bn}{\ensuremath{\mathrm{B^0}}\xspace}

\def\CP                {\ensuremath{\mathcal{CP}}\xspace}
\def\CPT               {\ensuremath{\mathcal{CPT}}\xspace} 
\def\C       {\ensuremath{\mathcal{C}}\xspace}

\def\T       {\ensuremath{\mathcal{T}}\xspace}

\newcommand{\kalpha}{\ensuremath{\mathrm{K_{\alpha}}}\xspace}
\newcommand{\kbeta}{\ensuremath{\mathrm{K_{\beta}}}\xspace}
\newcommand{\kfperp}{\ensuremath{\mathrm{K}}_{\nrightarrow f}\xspace}
\newcommand{\kfperpp}{\ensuremath{\mathrm{K}}^{\perp}_{\nrightarrow f}\xspace}
\newcommand{\kfperpone}{\ensuremath{\mathrm{K}}_{\nrightarrow f_1}\xspace}

\newcommand{\kfperponep}{\ensuremath{\mathrm{K}}^{\perp}_{\nrightarrow f_1}\xspace}
\newcommand{\kfperptwop}{\ensuremath{\mathrm{K}}^{\perp}_{\nrightarrow f_2}\xspace}

\long\def\symbolfootnote[#1]#2{\begingroup%
\def\thefootnote{\fnsymbol{footnote}}\footnote[#1]{#2}\endgroup}

\begin{document}

\title{
Can 
Future Observation of the Living Partner 
Post-tag
the Past Decayed State 
in Entangled Neutral K-Mesons ? 
}
\author{Jose Bernabeu}
\email{jose.bernabeu@uv.es}
\affiliation{Department of Theoretical Physics, University of Valencia, and
IFIC,~Univ.~Valencia-CSIC, E-46100 Burjassot, Valencia, Spain}
\author{Antonio Di Domenico}
\email{antonio.didomenico@infn.roma1.it}
\affiliation{Department of Physics, Sapienza University of Rome, and
INFN Sezione di Roma, P.le A.~Moro, 2, I-00185 Rome, Italy}


%
\begin{abstract}
Entangled neutral K-mesons allow the study of their correlated
dynamics at interference and decoherence times not accessible in any other system. We find novel quantum phenomena associated to a correlation-in-time between the two partners: the past state of the
first decayed kaon, when it was entangled before its decay, is 
post-tagged
 by the result and the time of the future observation of the second decay channel. This surprising ``from future to past'' effect 
is fully observable and
leads 
to the unique experimental tag of the \ksn-state,
an unsolved problem since the discovery of \CP violation.
\end{abstract}

\maketitle

\section{Introduction}
Since long ago, several authors have stated the crucial role that the neutral kaon system has played for understanding 
the intrincacies of the quantum world. In particular the words of R. Feynman~\cite{feynman}, T.D. Lee~\cite{lee} and L. B. Okun~\cite{okun} 
are all emphasising the uniqueness of this system as a jewel donated to us by nature.
They were referring to the peculiar properties of single neutral kaon states, which
display several rare phenomena like
the strangeness oscillation, the tiny mass splitting and the large difference in lifetimes of the physical states, the violation of the 
fundamental discrete symmetries
Charge-Parity (\CP) and Time-Reversal  (\T), the regeneration when traversing a slab of material.

  The present research is related to another peculiar character of neutral kaons: the ``strange entanglement", i.e. the entanglement which is specific to two neutral kaon systems with all the interconnections with the above properties.
  It is worth reminding here that the entanglement is one of the most striking feature of quantum mechanics, 
  as stressed by
  E. Schr\"odinger~\cite{schro}, in reply to the famous argument by A. Einstein, B. Podolsky and N. Rosen~\cite{epr} (EPR) based on local realism.
\par
Several tests of Quantum Mechanics and searches for possible decoherence and \CPT violation effects 
 that can exploit strange entanglement of neutral kaons 
 have
 been proposed~\cite{ref:HandbookAD,eberhard1,qmadd,selleri,uchi,bramonbell1,bramonbell2,bertbell1,genovese,bramonbell3,bramonbell4,hiesmayr1,hiesmayr2,nikitin1,qek1,qek2,shi1,shi2,eberhard,bertlmann1,bertlmann2,ellis1,ellis2,peskin,benatti1,benatti2,mavro1,mavro2,mavro3,kost1,kost2,kost3,kost4}.
The experimental investigation of strange entanglement started 
with the CPLEAR experiment~\cite{cplear}, 
and continued
with 
the KLOE and KLOE-2 experiments~\cite{kloe,kloe2,kloe2ws}
at 
DA$\Phi$NE~\cite{dafne1,dafne2,dafne3},
yielding several precision results 
\cite{kloeqm2006,kloeqm2010,kloeqm2021,kloecpt2013,kloe3pi,kloeas2,k2ichep2018,kloeksmu3,kloelifetime}. 
\par
The 
characteristic behaviour of strange entanglement,
with the peculiar properties of neutral
kaons not found in any other system, makes possible the exploration of
novel phenomena: the surviving correlation-in-time 
 from the 
observation of the future decay of the living partner at
      a given time to the identification
%
%
of the past kaon 
state leading to the first decay. This {\it from the future 
to the past} information
in a system with non-trivial time evolution, entering into times in which
the system was still entangled, could contribute to 
unveil the kind of reality to be associated to each part
of the system.

The methodology that we follow consists in  comparing
the description of the double decay distribution
at times $t_1$, $t_2$ with $\Delta t = t_2-t_1 > 0$ using (i) the formalism of the two decay times
state first introduced by Lee and Yang (LY)~\cite{leeyang,day,inglis,lipkin68}
with (ii)
the time history (TH) of the entangled state from  the 
coherent correlated neutral kaon system until its 
fate.
The quantum 
consistency of the two approaches and  the  $t_1$, $t_2$ symmetry
of the 
first
approach, with no special role of one of
the two decay times, naturally demand the study of a novel
problem: is it possible to infer the initial kaon state 
previous to the first decay
at $t_1$ from the observation of the second
decay at time $t_2 > t_1$,
i.e. 
a correlation able to provide
information from the future to the past? Contrary to the information 
from the past to the future, i.e., the prediction of the
kaon state at time $t_2$ from the observation of the 
first decay at time $t_1$, the question formulated in this
paper involves 
information on a part of the system at times in which the state was still fully entangled,
i.e. before the first decay,
when asking {\it which is which} is considered to be {\it unspeakable}
in John Bell's terminology~\cite{bell}. 
\par
In the following, first
we analyze the correlation
from past to future, i.e. which is the state before the second
decay at $t_2$, from the observation of the first decay channel
at $t_1$. Then
we infer the 
correlation
from
future to past, i.e. which is the state before the first decay 
at $t_1$, from the observation of the second decay channel at $t_2$.
We identify the decoherence region in $\Delta t$ in which the
surviving correlation tells us \kln at $t_2$ and \ksn at $t_1$, providing
the unique way to tag a \ksn experimentally.
We summarize the results presenting
some 
final 
remarks and our conclusion.
\section{From past to future}
\par
We 
consider an entangled two body neutral kaon system, as actually realised 
at 
DA$\Phi$NE, 
with $\phi\rightarrow \kn\knb$ decays, 
the source of EPR coherent $\kn\knb$ pairs
in the
$\C = -1$ antisymmetric state:
$| i \rangle 
= \frac{1}{\sqrt{2}} \{ |\kn \rangle |\knb \rangle - 
 |\knb \rangle |\kn \rangle \}$.
\par
Under particle exchange we call particle-1 the first one to decay at
time $t_1$, particle-2 the last to decay at time $t_2$. 
We remind the reader that Quantum Entanglement is associated to Non-Separability in two
aspects:
(i) we cannot identify which is which due to indistinguishability, and
(ii) we cannot specify the two parts of the system which  are not definite, showing 
that the parts have 
no local physical reality. 

\par
In fact, the antisymmetric state $|i\rangle$
is unique and 
therefore
 identically given in terms of any two generic linearly independent 
neutral kaon states,
 orthogonal or not~\cite{genericstate}.
%
As a particular case, it can be written
in terms
of the \ksn, \kln states with definite time evolution~\cite{Ksystem}:
$|i\rangle = \frac{\mathcal{N}}{\sqrt{2}} \{ |\ksn \rangle |\kln \rangle - 
 |\kln \rangle |\ksn \rangle \} $
with
$|\mathcal{N}|^2
=\left(1-|\langle\ksn|\kln\rangle|^2 \right)^{-1}
\simeq 1$~.
As a consequence,
the entangled state $| i \rangle$
at any time $t$
after its production remains 
unaltered,
even in presence of $\kn-\knb$ mixing:
\begin{widetext}
\begin{eqnarray}
|i (t) \rangle 
&=& \frac{\mathcal{N}}{\sqrt{2}} \{ |\ksn \rangle e^{-i\lambda_{S}t} |\kln \rangle e^{-i\lambda_{L}t}
-  |\kln \rangle e^{-i\lambda_{L}t} |\ksn \rangle e^{-i\lambda_{S}t} \} 
= 
e^{-i(\lambda_{S}+\lambda_{L})t} |i\rangle .
\label{eq:stateslgammatot}
\end{eqnarray}
\end{widetext}
If nothing is registered after the observation of the first decay at time $t_1$
(i.e. integrating over all subsequent decays at times $t_2$ of particle-2),
the {\it survival probability} of the entangled state 
is necessarily 
characterised
by the total width $\Gamma=\Gamma_S+\Gamma_L$ of the system~\cite{Csym}:
$P\left( t_1 \right) 
=\|~ | i (t=t_1) \rangle~\|^2 
= 
e^{-\Gamma t_1} $.
This also holds for any decay channel $t_1$-distribution with no
other subsequent observation.
%
%


\subsection{Two decay times state formalism (LY)} 
\par
Following the LY approach
of the 
two decay times 
entangled state (\ref{eq:stateslgammatot}),
the correlated state of the two partners decaying at times $t_1$ and $t_2$ can be formally written as~\cite{leeyang,day,inglis,lipkin68}:
\begin{widetext}
\begin{eqnarray}
|i_{t_1,t_2} \rangle &=& \frac{\mathcal{N}}{\sqrt{2}} \{ |\ksn \rangle e^{-i\lambda_{S}t_1} |\kln \rangle e^{-i\lambda_{L}t_2}
-  |\kln \rangle e^{-i\lambda_{L}t_1} |\ksn \rangle e^{-i\lambda_{S}t_2} \} .
\label{eq:stately}
\end{eqnarray}
\end{widetext}
The two decay times formalism defines in the combined
two terms of the entangled state  (\ref{eq:stateslgammatot}) what one calls particle-1 -- the first one to decay -- and particle-2 -- the second one
to decay. The (formal) use as evolution times
is justified because they are disjoint and there is 
no overlap between them: $t_1$ before, and
$t_2$ after, the performed
measurement
and its associated 
projection.
Accordingly, 
the decay amplitude of the initial state $| i \rangle$ to 
channel $f_1$
at time $t_1$ for particle-1 
and channel $f_2$ at time $t_2$ for particle-2,
and 
the corresponding observable double differential decay rate 
$I(f_1,t_1;f_2,t_2)$ 
can be readily calculated~\cite{buchanan,sanda,isidori,ref:HandbookAD}:
\begin{widetext}
\begin{eqnarray}
\label{eq:t1t2intensity}
  I(f_1,t_1;f_2,t_2)_{\rm LY}   
&=&
\left|  
\langle f_1(t_1) f_2(t_2) | T | i (t) \rangle 
\right|^2 
=
\left|  
\langle f_1 f_2 | T | i_{t_1,t_2} \rangle 
\right|^2 
\nonumber\\
&=&  C_{12} \{ |\eta_1|^2 e^{-\Gamma_L t_1 -\Gamma_S t_2}
+|\eta_2|^2 e^{-\Gamma_S t_1 -\Gamma_L t_2} 
-2 |\eta_1||\eta_2|e^{-{{(\Gamma_S+\Gamma_L)}\over{2}}(t_1+t_2)}\cos[ \Delta m 
\Delta t
+\phi_1-\phi_2]
\},
\end{eqnarray}
\end{widetext}
with 
$\langle f_i |T | \ksn \rangle$ and
$\langle f_i |T | \kln \rangle$ 
the decay amplitudes to the $f_i$ channel of \ksn and \kln,
$\eta_{i} \equiv |\eta_{i}| e^{i\phi_{i}}= \frac{\langle f_i |T | \kln \rangle}{\langle f_i |T | \ksn \rangle}~,$
and
$C_{12}={|\mathcal{N}|^2\over 2}| \langle f_1 |T|\ksn \rangle \langle f_2 |T|\ksn
\rangle |^2~.$
\par
As a corollary of the above approach 
one can notice that
at an intermediate step 
of the calculation
-- after the first decay at time $t_1$ --
the state of the surviving kaon (particle-2) immediately before its decay at time $t_2$ is expressed as:
\begin{widetext}
\begin{eqnarray}
|K^{(2)}  (t=t_2) \rangle
&=& \langle f_1 | T | i_{t_1,t_2}\rangle
=\frac{\mathcal{N}}{\sqrt{2}}  \langle f_1  |T | \ksn \rangle e^{-i(\lambda_S+\lambda_L) t_1}
\left[
e^{-i\lambda_L \Delta t}
|\kln \rangle -
\eta_1
 e^{-i\lambda_S \Delta t}
|\ksn \rangle
\right]~.
\label{eq:statek2ly}
\end{eqnarray}
\end{widetext}

Keeping $t_1$ and $f_1$ fixed -- the observation -- and renormalising the state at time $t_2=t_1$,
it corresponds to 
the evolution from time 
$t_1$ to time $t_2$
of the pure state
\begin{eqnarray}
|K^{(2)}(t=t_1) \rangle &=& 
\mathcal{N}_2 \left[
 |\kln \rangle -
\eta_{1}
 |\ksn \rangle
 \right]~,
\label{eq:statek2lyt1}
\end{eqnarray}
with $\mathcal{N}_2$ a suitable normalization factor. This is 
precisely the state of the living particle-2 
which cannot decay to $f_1$, as a result of the projection by the decay of particle-1 at $t_1$
as a filtering measurement
-- see eqs.(\ref{eq:kperp}) and (\ref{eq:stateff}) below.
\par
It is worth noting here that 
due to $\Delta \Gamma = \Gamma_S-\Gamma_L \neq 0$
two regimes can be identified in the time evolution of state (\ref{eq:statek2lyt1}):
(i) the generic interference region and (ii) the decoherence region, with
the relative weight of the \ksn component  negligible when
the following condition is satisfied:
\begin{eqnarray}
|\eta_{1}| e^{-\Delta \Gamma \Delta t / 2 } \ll 1 ~\hbox{~~~~~~~~~[\kln-tag]}.
\label{eq:conddtkl}
\end{eqnarray}
At long enough $\Delta t$ -- depending on what $f_1$ was -- the living partner is always a $|\kln\rangle$. 
This property is well understood and it has been used in the past in order to have \kln 
beams ``for all practical purposes" (FAPP) in Bell's terminology~\cite{bell}.

\subsection{Time history (TH)}
\par 
It is worth to point out that the result  (\ref{eq:statek2lyt1}) for
the living partner is in agreement with the EPR instantaneous
information 
due to the first decay when following the time history of strange entanglement, 
which we are now going to study in detail.
\par
We first notice that in the case of decay processes, any initial state has some probability per unit time 
to decay to a given decay 
channel $f$ except 
that 
with zero probability.
In particular the linear combination
\begin{eqnarray}
|\kfperp \rangle = \mathcal{N}_{\nrightarrow f} \left[ |\kln \rangle - \eta_f
 |\ksn \rangle \right]~,
\label{eq:kperp}
\end{eqnarray}
having a vanishing decay amplitude
$\langle f | T | \kfperp \rangle = 0$,
cannot decay to $f$.
This state is the one tagged for the unmeasured particle 
as a consequence of the projection imposed by the decay of the observed
particle. For the first decay to $f_1$ at time $t_1$, the tagged state of the surviving partner is given by Eq.(\ref{eq:kperp}) with $f = f_1$. 
In other words, the measured decay on one side prepares, 
in the quantum mechanical sense, its partner on the other side as a single kaon particle 
at a starting time $t = t_1$.
Then the $|\kfperp \rangle$ state freely evolves in time -- and in this sense the information is from past to future -- until its decay time at $t_2$, 
see  Eq.(\ref{eq:statek2ly}). We may ask whether this information constrains the past state of the decayed particle at $t_1$, which was undefined in the entangled system. This is a question that, for different scenarios, it is being debated in the literature -- see, for example Refs.\cite{post1,post2,post3,post4}. In our case, any state linearly independent to Eq.(\ref{eq:kperp}), orthogonal or not, leads to the same decay probability. This ``filtering identity''~\cite{filterid} is saying that the orthogonal component 
$| \kfperpp \rangle$
 is filtered from the past undefined state by the decay. The decay acts as a filtering measurement and, for calculation purposes, it is convenient to rewrite the entangled state at $t_1$, in terms of these two orthogonal states, as:
\begin{eqnarray}
| i \rangle 
&=& \frac{1}{\sqrt{2}} \{ | \kfperpp  \rangle |\kfperp \rangle - 
| \kfperp  \rangle |\kfperpp \rangle \}~.
\label{eq:stateff}
\end{eqnarray}
In this way, we may use the concept of transition probabilities at
the different relevant times in the history of the system.
\par
In summary 
four 
sequential steps are 
present
in the time history of the entangled state $| i \rangle$:
\begin{enumerate}
\item
the time evolution of the state $| i \rangle$ from time $t=0$ to time $t=t_1$, with definite total width $\Gamma$; 
\item
the projection 
of the state $| i (t=t_1) \rangle$ onto the orthogonal pair $| \kfperponep  \rangle |\kfperpone \rangle$,
filtered by the decay $f_1$, times
the decay amplitude of the state  $|\kfperponep \rangle$ into the $f_1$ channel; 
\item
the time evolution of the surviving (single) kaon state $|\kfperpone \rangle$ from time $t=t_1$ to time $t=t_2$;
\item
the projection at time $t=t_2$ of the evolved state $|\kfperpone (\Delta t)\rangle$ onto the state
$|\kfperptwop \rangle$ filtered by the decay $f_2$,
times the decay amplitude of the state  $|\kfperptwop \rangle$ into the $f_2$ channel.
\end{enumerate}
These steps 
straightforwardly 
lead
to 
the calculation of the observable double differential decay rate
by factorising the amplitudes as follows:
\begin{widetext}
\begin{eqnarray}
\label{eq:ddth}
  I(f_1,t_1;f_2,t_2)_{\rm TH}   
&=&
\left|  
\langle f_2 | T | \kfperptwop  \rangle
\langle \kfperptwop | \kfperpone(\Delta t) \rangle
\langle f_1 |T | \kfperponep  \rangle
\langle \kfperponep \kfperpone | i (t=t_1)\rangle
\right|^2~.
\end{eqnarray}
\end{widetext}
\par
One can easily verify that 
the TH approach is fully consistent with the LY approach~\cite{nogo}:
$I(f_1,t_1;f_2,t_2)_{\rm TH}   =   I(f_1,t_1;f_2,t_2)_{\rm LY}
 \equiv    I(f_1,t_1;f_2,t_2)$ .
%
\section{From future to past}
\par
As already pointed out,
the state (\ref{eq:statek2lyt1}) evaluated from
expression (\ref{eq:statek2ly}) 
in the LY approach coincides with the 
state $| \kfperpone \rangle$ of the surviving kaon after the first decay
in the TH approach. 
The $t_1$, $t_2$ symmetry of the correlated state in the LY approach -- Eq.(\ref{eq:stately}) --
with no special role of one of the
two decay times, 
demands the exploration of its implications when projecting it 
instead
onto the $f_2$ channel at time $t_2$.
With this information, the resulting past decayed state at time $t_1$ is:
\begin{widetext}
\begin{eqnarray}
\label{eq:statek1ly}
\boxed{
\begin{aligned}
|K^{(1)} (t=t_1) \rangle 
=& \langle f_2 | T | i_{t_1,t_2} \rangle
=&
\frac{\mathcal{N}}{\sqrt{2}}  \langle f_2 |T | \ksn \rangle 
\{
e^{-i\lambda_S t_1}
\left[\eta_2~
e^{-i\lambda_{L}t_2} 
|\ksn \rangle 
\right]-
e^{-i\lambda_L t_1}
\left[
e^{-i \lambda_S t_2}
|\kln \rangle
\right]
\}
~.
\end{aligned}
}
\end{eqnarray}
\end{widetext}
Expression  (\ref{eq:statek1ly}) 
corresponds to
the state of the decayed kaon (particle-1) immediately
before its decay at time $t_1$, once $t_2$ and $f_2$ are fixed 
for
the future ``fate'' of its partner.
Keeping $t_2$ and $f_2$ fixed -- the observation -- and varying the first decay time $t_1$, 
it
corresponds to the single kaon evolved state, before the first decay,
from time $t=0$ to time $t=t_1$ of the state
\begin{eqnarray}
|K^{(1)}(t=0) \rangle =
\mathcal{N}_1 \{ & &
\eta_{2}
e^{-i\lambda_L t_2}
 |\ksn \rangle 
-e^{-i \lambda_S t_2}
 |\kln \rangle
 \}~
\label{eq:statek1lyt1}
\end{eqnarray}
with $\mathcal{N}_1$ a suitable renormalization factor.
{\bf
Contrary to eq.(\ref{eq:statek2lyt1}) which is independent on the past $t_1$ decay time, 
eq.(\ref{eq:statek1lyt1}) shows a dependence not only on the decay channel $f_2$, but
also on the future $t_2$ decay time.
}
\par
This is a striking result which clearly involves 
a correlation-in-time
from the future
observation at time $t_2$ to the 
past,
inferring the initial
kaon state {\it before} its first decay at $t_1$.
It becomes well defined during
the time evolution of the 
entangled state $|i\rangle$ described by eq.(\ref{eq:stateslgammatot}) when the state of particle-1 (and particle-2)
should have been undefined
in the absence of any observation.
We insist that the 
{\it post-tagging}
implied by Eq.(\ref{eq:statek1ly})
is not an artefact of the formalism  but a factual observable
accessible to experimental studies and thus it is fully physical.
In a time history from future to past, the
future observation at time $t_2$ tags particle-1 at the time $t_1 = t_2$
into the state  proportional to  
$\{ \eta_{2} |\ksn \rangle - |\kln \rangle \}$,
%
the state not decaying to $f_2$. Keeping
$t_2$ and $\eta_2$ fixed -- the observation --, 
the backward evolution of this tagged 
unobserved
state 
to $t_1 < t_2$ leads to Eq.(\ref{eq:statek1ly}).

\subsection{The interference and decoherence regimes: the \ksn tag}
\par
\begin{figure}
   \centering
 \includegraphics[width=3.4in]{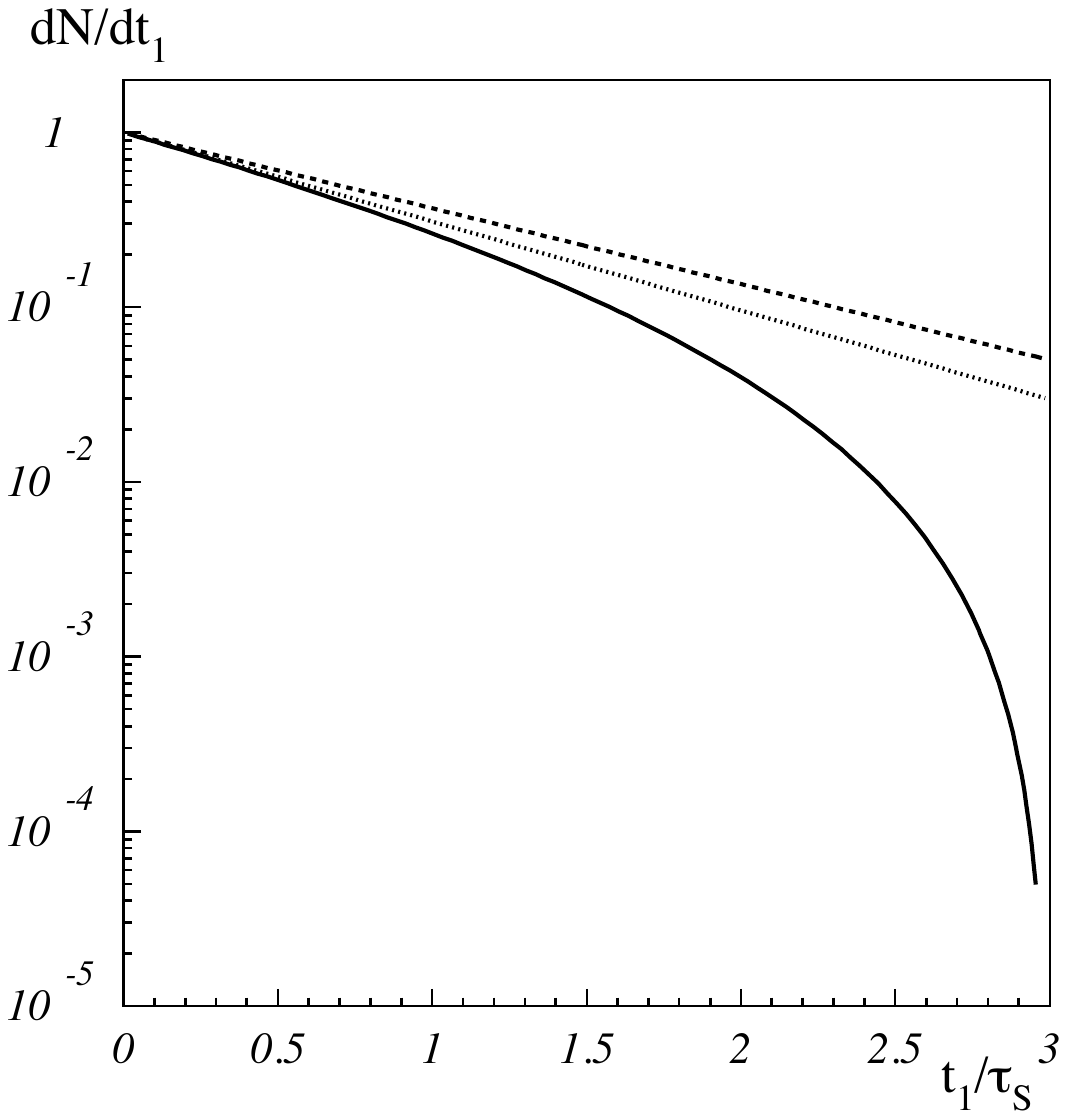} 
   \caption{
   The decay rate distribution into a generic channel $f_1$ 
      of state (\ref{eq:statek1ly})
as a function of $t_1$ 
for the future observation at $t_2=3~\tau_S$ 
(solid line),
and when condition (\ref{eq:conddtks}) for decoherence is
satisfied (dashed line), 
with $f_2=f_1$.
The last shows a definite lifetime $\tau_S$ and does not depend on the decay channel $f_1$.
They differ both from the $t_1$-distribution
                                     (dotted line) in the absence of a future measurement 
                                     (in this case $\Gamma_L$ has been multiplied by a factor 100 to appreciate graphically the difference between dotted and dashed lines).
                                     All
distributions are normalised to unity at $t_1=0$.
 }
   \label{fig1}
\end{figure}

As a counterpart of the observability of the {\it pre-dicted} Eq. (\ref{eq:statek2ly})
through the $t_2$ time distribution of the second decay, once the first decay to the $f_1$ decay channel at $t_1$ 
is fixed, the $t_1$ time distribution of the
first decay as
{\it post-dicted} in Eq. (\ref{eq:statek1ly}) is also observable, once the second decay channel $f_2$ and the decay time $t_2$ are fixed. 
As function of $t_1$, two
different regimes can be identified: the generic interference region,
in which the $t_2$ dependence of 
Eq. (\ref{eq:statek1ly})
is apparent,
and the decoherence region, in which 
the relative weight of the
\kln component is negligible. Decoherence is reached for 
large
$\Delta t$ satisfying the condition:
\begin{eqnarray}
\boxed{
\begin{aligned}
 e^{-\Delta \Gamma \Delta t / 2 } / |\eta_{2}| \ll 1~\hbox{~~~~~~~~~[\ksn-tag]}~,
\end{aligned}
}
\label{eq:conddtks}
\end{eqnarray}
leading to
a pure \ksn beam before the first decay.
This consequence of the surviving correlation-in-time is most
rewarding.
Due to \CP violation and the non-orthogonality of the stationary states $\langle\kln|\ksn\rangle\neq0$, 
there is no decay channel able to tag either \ksn or \kln on an event-by-event basis.
While it is relatively easy to prepare FAPP pure \kln beams,
fulfilment of condition (\ref{eq:conddtks}) constitutes the only known FAPP method to 
actually {\it post-pare}
a \ksn beam 
(i.e. the short-lived 
stationary state)
with arbitrary high purity (depending on $\Delta t$ and $\eta_2$),
preparation otherwise impossible with other methods.
As 
illustration
of the observables in the two different regimes, Figure~\ref{fig1}
shows 
the decay rate distribution into 
a generic 
channel $f_1$ 
of state (\ref{eq:statek1ly})
as a function of $t_1$ in two cases: 
either observed at $t_2=3~\tau_S$ (interference region) or when condition (\ref{eq:conddtks}) is satisfied (decoherence region),
with $f_2=f_1$ to maximize the interference effects and make visible the difference between the two cases. 
This choice $f_2 = f_1$  also emphasizes the differing results as
due to the dependence on the time of the future observation.
{\bf Whereas the decoherence case 
shows a definite width $\Gamma_S$, 
the future observation in the interference region leads to a $t_1$-distribution with no definite lifetime.}
In the latter case
the $t_1$ distribution does depend on the decay channel.
All these results differ from the
time distribution, given by the total width $\Gamma$, in the absence of 
any future observation~\cite{survivalprob,kloeprog}.


\section{Remarks and Conclusions}
\par
    In the case of entangled neutral mesons in the $\C =(-)$ state, the dynamics before the first decay was considered to be trivial, even with mixing, as corresponding to a definite time evolution with the total width of the system. 
    Hence in the past the experimental studies 
    were
    concentrated in the 
   observation of the single kaon decay rate distribution 
    between the two decays depending on
$\Delta t$. The first decay acts as a filtering preparation of the {\it initial} single state of the living partner.
     Our paper demonstrates the consistency of this description in
terms of observables 
like: {\it What is the state of the living
partner at the time $t_2$ of the second decay?}, a well defined -- speakable --
question,
because at 
$t_2$ the system is no longer 
entangled,
 after the measurement of the first decay channel $f_1$ at $t_1$.
Asking {\it What was the state of the first decayed kaon?}, however, was considered to be unspeakable,
because before the first decay the
system was entangled. 
Our study from future to past leads to
the conclusion that this last contention is only valid as long as no future observation is made.
{
In our case of a {\bf future} measurement, the independence on the reference frame is assured when
a time-like interval between the two decays is considered.
 }
\par
     The present research has gone indeed a step further and seems to recognize that the correlation between the two partners survives their explicit dynamics, with a transition from the quantum correlation of entanglement to a classical correlation of separable \ksn \kln
     states. The     
     information 
     from a 
     measurement on the living partner -- $f_2$ decay channel at $t_2$ -- 
     to the state of the decayed meson at $t_1$ 
     is most surprising. Entering into the entangled region, i.e. just before the first decay, is not an artefact of our formalism, but a precise experimental observable through the $t_1$ distribution of decays in any channel,
     as shown in Fig.\ref{fig1}. 
     This is so for even situations of decoherence, 
     Eq.(\ref{eq:conddtks}),
     a physical situation only reachable for strange entanglement with the two very different lifetimes of 
     \kln and \ksn. The relevance of this result for particle physics is outstanding: 
     the unique way to tag {\it what a \ksn is}, i.e.
     the solution of an open problem 
     since the discovery of \CP violation.
\par Our results seem to confirm the counter-intuitive feature of
time in quantum mechanics. The surviving correlation-in-time found here goes beyond other
phenomena, like delayed choice experiments, quantum erasers or teleportation, discussed
for 
photons~\cite{qe1,qe2,qe3,bert1,zeilrev}.
In the case of delayed choice designs, the system is stationary at all times
and the choice of
the outcome can be made by either advanced or delayed observation with
the result unchanged~\cite{qeknote}.
In the effect discussed here, 
the tagged state of the past decayed kaon has a non-trivial time dependence with a result depending on the
decay time of the future observation ({post-tagging}),
as depicted in Fig.~\ref{fig1}.
%
The result is not symmetric when comparing the outcomes in the two senses of 
``from past to future", Eq.(\ref{eq:statek2ly}), and
``from future
to past",
Eq~(\ref{eq:statek1ly}).
This is 
characteristic of
the neutral K-meson system with flavor mixing, $\Delta\Gamma>0$ and $\langle\kln|\ksn\rangle\neq0$,
leading
at decoherence
times 
to the unique experimental \ksn-tag .
These predictions are fully observable through the measurement of $t_1$-distributions as the ones shown 
in Fig.~\ref{fig1}, with no analogue in other physical systems.

\par 
Our results demonstrate that the correlation-in-time is definite between the outcome at a given time of the observed decay and the state of the unobserved partner. This correlation also survives when the observation is made in the future, when the system is no longer entangled after the first decay, post-tagging the past state of the unobserved decayed partner depending on the result and the time of
the future observation. The non-trivial time evolution for the neutral K-meson system leads to a result which is non-symmetric in time. As a consequence, it opens the way to a novel kind of experimental studies, not envisaged before.
%
\begin{acknowledgements}
 This research has been supported by the FEDER/MCIyU-AEI Grant FPA2017-84543-P and Generalitat Valenciana Project GV PROMETEO 2017-033 .
\end{acknowledgements}


\begin{thebibliography}{1}
\bibitem{genericstate}
In fact, 
the antisymmetric state $| i \rangle$ can be 
expressed as:
$| i \rangle =\frac{\mathcal{N}_{\alpha\beta}}{\sqrt{2}} \{ | \kalpha\rangle | \kbeta \rangle - 
 | \kbeta\rangle | \kalpha\rangle \}$,
with  
$|\kalpha\rangle$ and $| \kbeta \rangle$
two 
arbitrary
linearly 
independent 
superpositions of $| \kn \rangle$ and
$| \knb \rangle$ states, and 
$|\mathcal{N}_{\alpha\beta} |^2=\left(1-|\langle\kalpha|\kbeta\rangle|^2 \right)^{-1}$.
\bibitem{Ksystem}
The eigenstates of the effective Hamiltonian describing a single neutral kaon are
the short- and long-lived 
physical states $|\ksn\rangle$ and $|\kln\rangle$.
They have definite masses $m_{S,L}$ and lifetimes $\tau_{S,L}$ and
evolve as a function of the kaon proper time $t$ as pure exponentials 
$|\ksln(t)\rangle=e^{-i\lambda_{S,L}t}|\ksln\rangle$
with $\lambda_{S,L}=m_{S,L}-i\Gamma_{S,L}/2$, and $\Gamma_{S,L}=(\tau_{S,L})^{-1}$.
They are usually expressed in terms of the flavor 
eigenstates $|\kn\rangle$, $|\knb\rangle$ as:
$|\ksln\rangle \propto
\left[ (1+\epsilon_{S,L}) |\kn \rangle
\pm (1-\epsilon_{S,L}) |\knb \rangle
\right]$
with $\epsilon_S$
and $\epsilon_L$
two small complex parameters describing the \CP impurity in the physical states,
making them non-orthogonal, $\langle\ksn|\kln\rangle\simeq\epsilon_L+\epsilon_S^{\star}$.
One can equivalently define
${\epsilon} \equiv (\epsilon_S+\epsilon_L)/2$, and
$ \delta \equiv (\epsilon_S-\epsilon_L)/2$;
adopting a suitable phase convention 
$\epsilon\neq0$ implies \T violation,
$\delta\neq0$ implies \CPT violation,
while $\delta\neq0$ or
$\epsilon\neq0$ implies \CP violation.
{It is worth noting that one of 
the most stringent tests of the fundamental \CPT symmetry is performed
comparing the measured \CP violation in \ksn and \kln states~\cite{ref:HandbookAD,kloeas2}~.}
\bibitem{Csym}
This property holds only for the $\C = -1$ antisymmetric state, but not for the $\C = +1$ symmetric state 
in which the time evolution would induce $\kn\kn$ and $\knb\knb$ terms due to $\kn-\knb$ 
mixing by weak interactions.
\bibitem{nogo}
It is worth mentioning that the concepts of quantum entanglement and decay as a filtering measurement -- as used in the TH approach -- 
have been instrumental 
in recent years~\cite{banuls1,banuls2,btviol} for observing  \T-Violation for the \bn-meson~\cite{babartviol},
 bypassing the well known
NO-GO argument for measuring genuine \T and \CPT asymmetries in transitions for decaying particles.  
Similar ideas have been formulated for the case of neutral kaons~\cite{ktviol,kcptviol}.
\bibitem{survivalprob}{
Even though it would be more challenging experimentally, also the survival probability of state (\ref{eq:statek1ly}),
$P^{(1)}(t_1)=\|~ |K^{(1)}(t=t_1) \rangle~\|^2 $,
could be measured as a function of $t_1$, which is also a characteristic of the decaying state and not related to any particular choice of the final state $f_1$.
}
\bibitem{qeknote}{
The same conclusion holds for the proposed 
``kaonic'' version of the quantum eraser~\cite{qek1,qek2,qe3}, in which the time 
of the observation of the meter 
is 
irrelevant for the searched effect,
whether or not 
in the delayed choice mode.
}
\end{thebibliography}
\end{document}